\documentclass[twocolumn,amsmath,amssymb,aps,pra]{revtex4-1}
\usepackage{float}
\floatstyle{boxed} 
\newcommand{\bra}[1]{\ensuremath{\left\langle#1\right|}}
\newcommand{\ket}[1]{\ensuremath{\left|#1\right\rangle}}
\newcommand{\bracket}[2]{\ensuremath{\left\langle#1 \vphantom{#2}\middle|  #2 \vphantom{#1}\right\rangle}}

\newcommand{\comb}[2]{\ensuremath{\left(\begin{array}{c}#1\\#2\end{array}\right)}}
\usepackage{color}
\usepackage{graphicx}
\usepackage{dcolumn}
\usepackage{bm}
\usepackage[utf8]{inputenc}
\usepackage[T1]{fontenc}
\usepackage{hyperref}

\begin{document}

\title{Requirements for two-source entanglement concentration} 

\author{Mihai Vidrighin$^{1,2}$, Tim J. Bartley$^{1}$, Gaia Donati$^{1}$, Xian-Min Jin$^{1,3}$, Marco Barbieri$^{1}$, W. Steve Kolthammer$^{1}$, Animesh Datta$^{1}$, and Ian A. Walmsley$^{1}$ }
\affiliation{$^1$Clarendon Laboratory, Department of Physics, University of Oxford, OX1 3PU, United Kingdom}
\affiliation{$^2$QOLS, Blackett Laboratory, Imperial College London, London SW7 2BW, United Kingdom}
\affiliation{$^3$Department of Physics, Shanghai Jiao Tong University, Shanghai 200240, PR China}

\begin{abstract}
There have been several experimental investigations of entanglement enhancement of a two-mode squeezed vacuum. In particular, conditional preparation by photon subtraction has been shown to improve correlations achieved with this entangled resource. Here we analyse the role of Gaussian and non-Gaussian measurement for entanglement concentration acting on a pair of two-mode squeezed states. We find stringent requirements for achieving further entanglement enhancement by a joint measurement setup on the two resources.
\end{abstract}

\maketitle

\section{Introduction}
In recent years, quantum optics experiments have demonstrated the possibility of quantum enhanced communication by coherent manipulation of states of light. Important and illustrative quantum communication tasks such as quantum cryptography \cite{crip}, quantum dense coding \cite{dense} and quantum teleportation \cite{tele} can be implemented within the continuous variables (CV) framework,  by manipulating the quadrature amplitudes of electromagnetic modes, specifically in Gaussian states \cite{review, paris, newcv}. This type of implementation is readily achieved in the laboratory, relying on linear optics for deterministic transformations, and on the nonlinear process of parametric down-conversion (PDC) for generation  of non-classical and entangled squeezed Gaussian states. A challenge for this approach is that the fidelity of CV quantum communication depends on the quality of entanglement and unit fidelities are only achieved by employing infinitely squeezed states which are impossible to realise in practice. In addition, quantum transmission channels degrade shared entanglement, decreasing the fidelity of communication protocols.

In order to improve the performance of CV quantum communication protocols, methods have been developed to enhance entanglement initially shared in the form of bipartite Gaussian states. The term \textit{entanglement distillation} \cite{popescu, concentration} serves as a label for protocols consisting of local operations and classical communication that convert a number of identical two-mode entangled states into a smaller number of more entangled two-mode states. More specifically, when the initial states are pure, the protocols are termed entanglement concentration. Entanglement distillation cannot be achieved if both the initial states and the local operations are Gaussian \cite{plenio, cirac, fiurasec}.

In practice, a non-Gaussian operation may be achieved by measurement with single-photon detectors \cite{apds}. Photon subtraction and photon addition on the modes of an entangled two-mode Gaussian state have both been shown to increase entanglement. These operations have been extensively discussed in recent experimental \cite{nii, exp2} and theoretical \cite{kim} reviews. In the context of entanglement concentration, both multiple applications \cite{mit} and the coherent superposition \cite{chinezi} of annihilation and creation operators -- well-approximated by photon subtraction and addition -- have been considered. It has been shown that photon subtraction can improve the fidelity of teleportation with CV, while simple photon addition cannot \cite{oliva1, ilu0, chinez}. Photon-subtracted states have been shown to be more tolerant to loss than Gaussian states \cite{ilu}. The enhancement of photon subtraction on the nonlocal behaivour of two-mode squeezed states has also received attention \cite{oliva2, oliva3}.

Experimentally, an enhancement of entanglement in an initial two-mode Gaussian state has been demonstrated by various methods: photon subtraction in either or both modes \cite{taka}, non-local photon subtraction \cite{oju}, and weak Gaussian measurement after the state has gone through a non-Gaussian noise channel \cite{mixed}.

In this work we compare different measurement operations acting jointly on a pair of two-mode Gaussian entangled states with the aim of performing entanglement concentration. In particular, we explore the advantage of pairwise interaction of resource states in the first step of the iterative protocol introduced in \cite{plenio2, plenio3}, employing photon subtraction as the entanglement-enhancing non-Gaussian operation. The approach consists of combining two photon subtracted states on beam splitters and performing a Gaussian measurement on two of the output modes. It represents an example of two-source entanglement distillation feasible within the current experimental state of the art. 

We find analytical expressions for the result of interfering different photon subtracted states. We construct the best Gaussian measurements required to combine the entangled states and identify the arrangement for achieving the optimal trade-off between success probability and enhancement of the entanglement. Finally, we analyse the resilience to loss of the approach. Our results show that, if limited to the first step of the iterative distillation protocol, combining photon-subtracted states has limited advantages with respect to simple photon subtraction.

\section{Entanglement concentration by photon subtraction on a two-mode squeezed vacuum}
A two-mode squeezed vacuum (TMSV) is commonly produced by spontaneous non-degenerate PDC, a process occurring when a noncentrosymmetric crystal is pumped by an intense laser field. A photon can be spontaneously annihilated in the pump field and one photon created in each of the two modes $a$ and $b$, resulting in the state
\begin{equation}
\ket{\psi_{TMSV}}=\sqrt{1-\lambda^2}\sum_n\lambda^n\ket{n}_{a}\ket{n}_{b}
\label{eq:1}
\end{equation}
where $\ket{n}_c$ is a number state with $n$ photons in mode $c$ and $\lambda=\tanh(r)$ is determined by the parameter $r$ that is proportional to both the nonlinear coefficient of the PDC crystal and the intensity of the pump laser beam. A TMSV is the typical entangled resource for CV quantum communication protocols.\\
Photon subtraction acting on a TMSV provides a probabilistic increase in entanglement for which a trade-off exists between the \textit{probability of success} and the \textit{entanglement of the output state}. In the following, we quantify entanglement according to the entanglement negativity \cite{vidal}, a monotone defined as
\begin{equation}
\mathcal{N}(\rho_{AB})=\frac{||\rho^{T_B}||-1}{2},
\end{equation}
where $||\sigma||=\text{Tr}[\sqrt{\sigma^{\dagger}\sigma}]$, and $\rho^{T_B}$ is the partially transposed density matrix of the bipartite system.
\begin{figure}[H]
\centering
\includegraphics[width=0.2\textwidth]{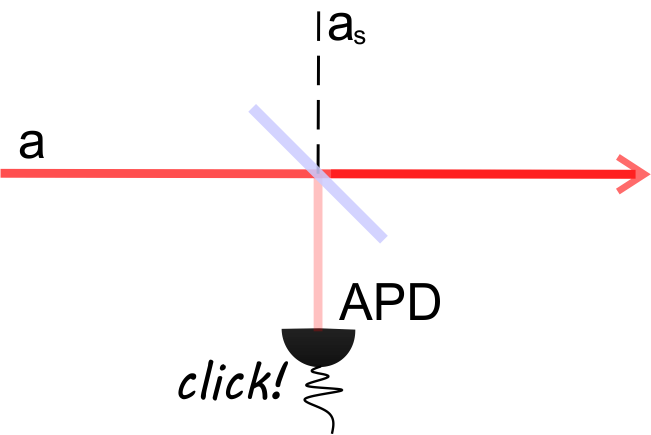}
\caption{Implementation of photon subtraction: photons are subtracted from the input mode $a$ by means of a high transmittivity beam splitter which reflects a small fraction of the input intensity towards a photon-number resolving detector (PNRD). The PNRD detecting an event heralds successful photon subtraction\cite{knill}. An approximate  PNRD can be conveniently implemented by multiplexed detection with avalanche photodiodes (APDs) \cite{grangier, taka5}}
\label{fig:1}
\end{figure}

\begin{figure}[H]
\centering
\includegraphics[width=0.45\textwidth]{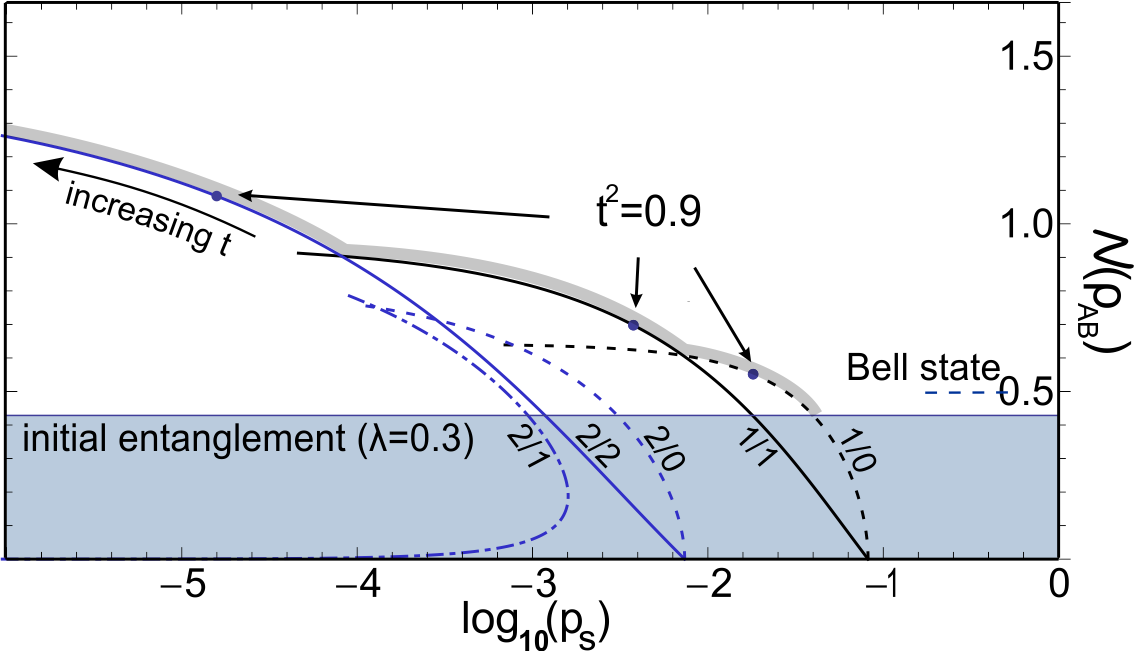}
\caption{Negativity of entanglement $\mathcal{N}(\rho_{AB})$ as a function of the logarithm of the success probability $p_s$, for different photon subtraction strategies. These quantities are reported as a parametric plot when the transmittivity $t$ of the subtracting beam splitter is varied. The curves are labeled $k_A/k_B$ where $k_A$($k_B)$  is the number of photons subtracted from the mode A(B). For reference, the negativity of entanglement of some subtracted states for $t^2=0.9$ as well as a Bell state are marked. The optimum subtraction strategy is highlighted in gray.} 
\label{fig:2}
\end{figure}

In order to find the measurement operator corresponding to photon subtraction as depicted in Figure~\ref{fig:1}, we use the expansion of the beam splitter operator \cite{theory}:
\begin{equation}
\hat{B}_{aa_s}(\theta)=e^{-\tan\frac{\theta}{2}\hat{a}\hat{a}_s^{\dagger}}e^{-\ln(\cos\frac{\theta}{2})(\hat{a}_s^{\dagger}\hat{a}_s-\hat{a}^{\dagger}\hat{a})}e^{\tan\frac{\theta}{2}\hat{a}^{\dagger}\hat{a}_s}
\label{eq1}
\end{equation}
where, $\cos(\frac{\theta}{2})=t$ is the transmittivity of the beam splitter and $\sin(\frac{\theta}{2})=r$ its reflectivity. The ancillary mode $a_s$, initially in the vacuum state, interacts with mode $a$ on the beam splitter and is then projected out. Using Eq.~\ref{eq1}, the measurement operator corresponding to the subtraction of k-photons from mode $a$ is $\hat{M}_{k,a}$ 
\begin{equation}
\begin{split}
\hat{M}_{k,a}&=_{a_s}\!\bra{k}\hat{B}(\theta)_{aa_s}\ket{0}_{a_s}=\frac{(-\tan\frac{\theta}{2})^k}{\sqrt{k!}}\hat{a}^k t^{\hat{a}^{\dagger}\hat{a}}\\&=\frac{1}{\sqrt{k!}}(-\frac{r}{t})^k\hat{a}^k t^{\hat{n}}.
\end{split}
\end{equation}
The unnormalized state created by subtracting one photon from a TMSV is found to be
\begin{equation}
\ket{\psi_{1,0}}=\sqrt{1-\lambda^2}r\lambda\sum_{n,m}(\lambda t)^n\sqrt{n+1}\ket{n}_a\ket{n+1}_b
\end{equation}
by acting $\hat{M}_{1,a}$ on $\ket{\psi_{TMSV}}$.
The norm squared of this state gives the probability of photon subtraction $p_s=\bracket{\psi_{1,a}}{\psi_{1,a}}$. Increasing the transmittivity $t$ of the beam splitter used for photon subtraction increases the negativity of entanglement of this state but decreases the probability of photon subtraction. This trade-off appears when considering different subtraction strategies, as shown in Figure~\ref{fig:2}. 
Notably, different strategies are advantageous in different regimes of entanglement concentration. The optimal trade-off, highlighted in gray, will serve as a reference to analyse the efficiency of the schemes considered in the following section.
\section{Combining Two Gaussian states}
\begin{figure}
\centering
\includegraphics[width=0.4\textwidth]{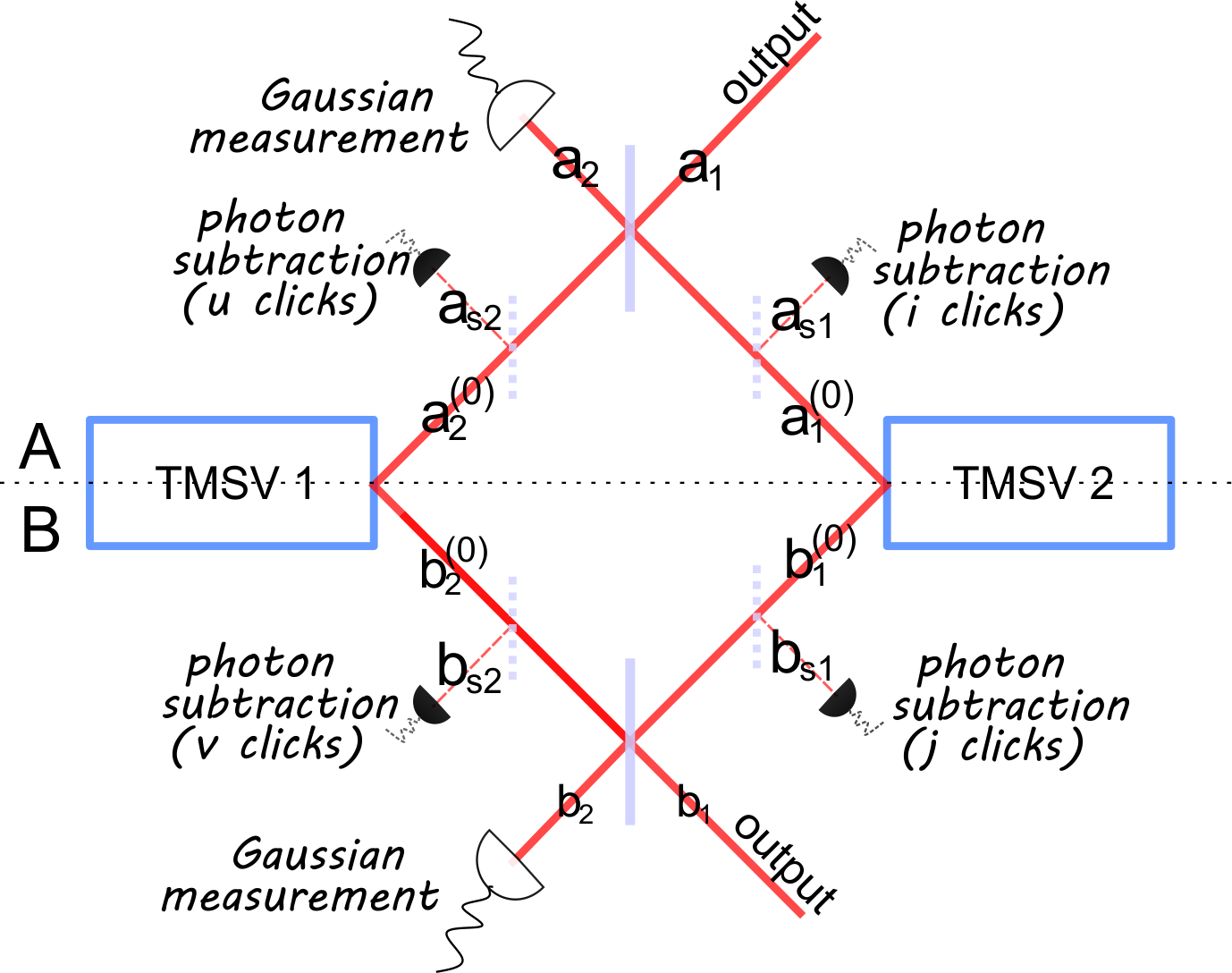}
\caption{First step in the iterative protocol proposed in \cite{plenio2}. A and B share two TMSVs de-Gaussified by photon subtraction. The states are combined locally on a 50:50 beam splitter. A Gaussian measurement leaves two of the modes in states with increased entanglement.} 
\label{fig:3}
\end{figure}
Performing CV entanglement distillation from Gaussian states and ending with Gaussian states requires de-Gaussification of a large number of resources followed by an iterative protocol that reproduces a Gaussian state \cite{plenio2}. One Gaussification step consists of the combination of pairs of identical states, employing 50:50 beam splitters and vacuum projection. This linear-optical scheme enhances entanglement as the outputs converge towards a Gaussian state. While the resources required for such a protocol increase rapidly with the number of iterations \cite{josh}, the first step alone provides a feasible setup for distillation starting with a pair of two-mode states.
We now give analytic expressions for the four-mode states obtained by combining different photon-subtracted states on beam splitters as pictured in Figure~\ref{fig:3}.
Throughout this section, the results of non-trace-preserving operations are given as unnormed Hilbert space vectors whose  norm squared equals the probability of successful implementation. The result of subtracting $i$ and $j$ photons, respectively, from the modes of a TMSV, will be denoted $\ket{\psi_{i,j}}$.
The result of subtracting $i$, $j$, $u$ and $v$ from modes $a_1$, $a_2$, $b_1$ and $b_2$, respectively, as shown in Figure~\ref{fig:3}, and combining the two photon-subtracted states on beam splitters, will be denoted $\ket{\Psi_{i,j,u,v}}$. In order to obtain closed form expressions, we need to presenve the symmetry of the states as much as possible, hence consider the same number of beam splitters must be inserted in the modes of states 1 and 2. Thus, when necessary, we consider a beam splitter inserted on one side but no photons detected in the reflected mode. We denote these events with the symbol $\tilde 0$. The results for various one-photon subtraction schemes are
\begin{widetext}
\begin{equation}
\begin{split}
\ket{\Psi_{1,0,\tilde{0},0}}&=\frac{1}{\sqrt{2}}(\ket{\psi_{1,0}}_{a_1,b_1}\otimes\ket{\psi_{\tilde{0},0}}_{a_2,b_2}+\ket{\psi_{\tilde{0},0}}_{a_1,b_1}\otimes\ket{\psi_{1,0}}_{a_2,b_2};\\
\ket{\Psi_{1,1,\tilde{0},\tilde{0}}}&=\frac{1}{2}[(1-\lambda^2)r^2\lambda\sum_{n,m}(\lambda t^2)^{n+m}(n+m+2)\ket{n,n}_{a_1,b_1}\otimes\ket{m,m}_{a_2,b_2}+(\ket{\psi_{1,\tilde{0}}}_{a_1,b_1}\otimes\ket{\psi_{\tilde{0},1}}_{a_2,b_2}+\ket{\psi_{\tilde{0},1}}_{a_1,b_1}\otimes\ket{\psi_{1,\tilde{0}}}_{a_2,b_2})];\\
\ket{\Psi_{1,0,0,1}}&=\frac{1}{2}[\frac{(1-\lambda^2)r^2\lambda}{t}\sum_{n,m}(\lambda t)^{n+m}(n-m)\ket{n,n}_{a_1,b_1}\otimes\ket{m,m}_{a_2,b_2}+(\ket{\psi_{1,0}}_{a_1,b_1}\otimes\ket{\psi_{0,1}}_{a_2,b_2}-\ket{\psi_{0,1}}_{a_1,b_1}\otimes\ket{\psi_{1,0}}_{a_2,b_2})];\\
\ket{\Psi_{1,0,1,0}}&=\frac{1}{\sqrt{2}}(\ket{\psi_{2,0}}_{a_1,b_1}\otimes\ket{\psi_{\tilde{0},0}}_{a_2,b_2}+\ket{\psi_{\tilde{0},0}}_{a_1,b_1}\otimes\ket{\psi_{2,0}}_{a_2,b_2})\\
\ket{\Psi_{1,1,1,1}}&=\frac{1}{2}[\frac{(1-\lambda^2)r^4\lambda^2}{2}\sum_{n,m}(\lambda t^2)^{n+m}((n+1)(n+2)+(m+1)(m+2))\ket{n,n}_{a_1,b_1}\otimes\ket{m,m}_{a_2,b_2}+\\
&+(\ket{\psi_{2,\tilde{0}}}_{a_1,b_1}\otimes\ket{\psi_{\tilde{0},2}}_{a_2,b_2}+\ket{\psi_{\tilde{0},2}}_{a_1,b_1}\otimes\ket{\psi_{2,\tilde{0}}}_{a_2,b_2})].\\
\end{split}
\label{mare}
\end{equation}
\end{widetext}
The method employed to derive the expressions is described in the Appendix.
We now discuss the Gaussian measurement in Figure~\ref{fig:3}. In the original proposal \cite{plenio2}, vacuum projection was used. Here, we look at a more general scenario, with Gaussian measurements implemented by balanced homodyne detection (i.e. a projection on quadrature eigenstates) and 8-port homodyne detection (i.e. projection on coherent states). The use of the latter setup for vacuum projection has been discussed previously \cite{plenio3, feito}.
 
The phase space distribution of a Gaussian state is completely described by the displacement vector (mean position in phase space) and the covariance matrix of the quadrature amplitudes. The covariance matrix of states conditioned on the result of a Gaussian measurements on Gaussian states does not depend on the particular measurement result. Only the displacement of the obtained state is a function of the measurement result. Therefore, given the result and the initial state, the output can always be centered to the origin by appropriate post-measurement displacements \cite{cirac, fiurasec}. Therefore if the Gaussian measurement is performed before the single photon detections, for any given outcome the state corresponding to a null displacement vector can be obtained by classical communication and local displacement of modes $a_1, b_1, a_{s1}, a_{s2}, b_{s1},\text{ and } b_{s2}$, as denoted in Figure~\ref{fig:3}. Since the entanglement properties of states depend only on their covariance matrix, we can assume that the Gaussian projections in the setup of Figure~\ref{fig:3} are implemented deterministically.

\begin{figure}[H]
\centering
\includegraphics[width=0.42\textwidth]{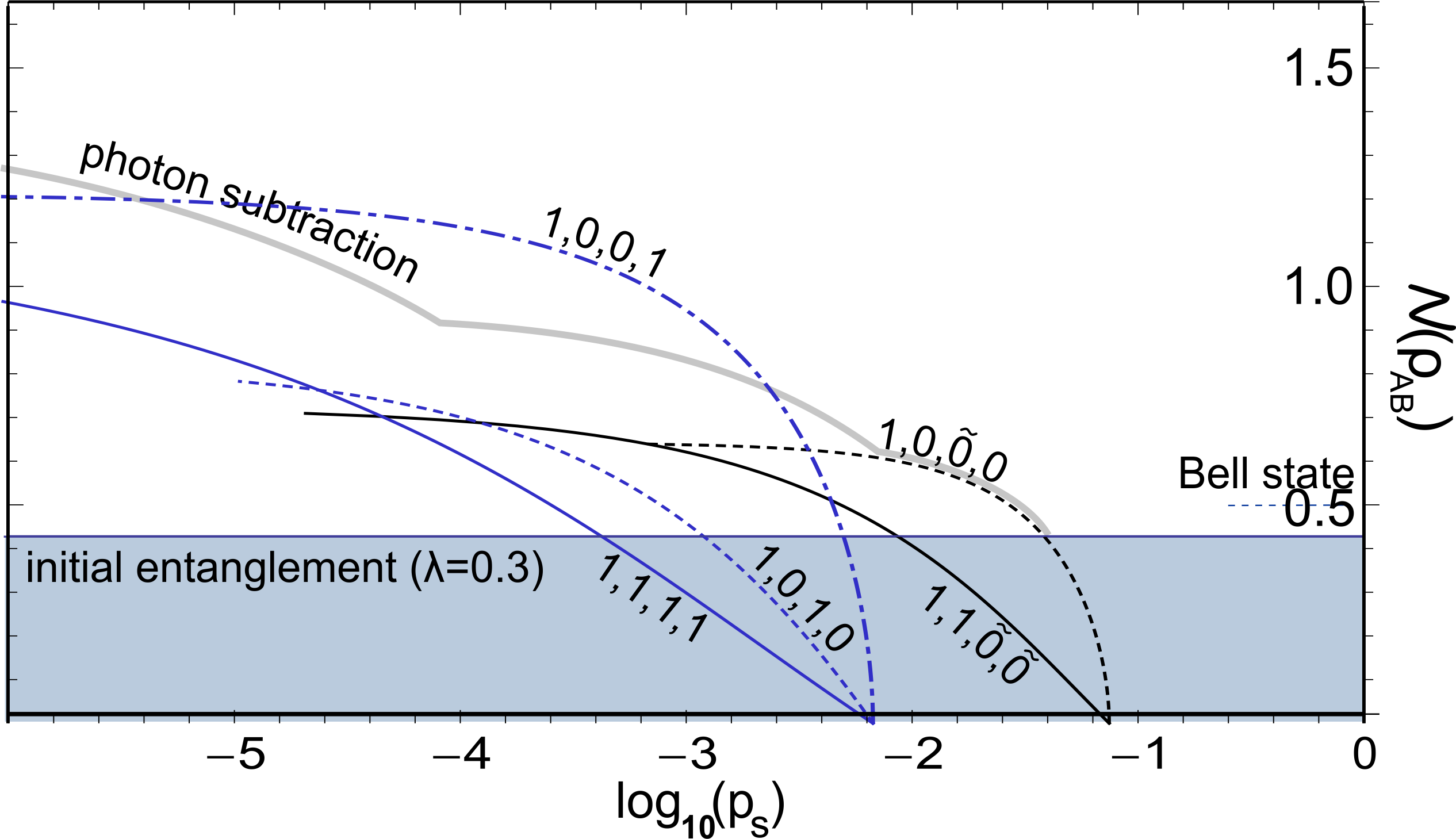}
\label{mat}
\caption{Negativity of entanglement versus the logarithm of the success probability for different joint measurement strategies, yielding states given in Eq. \ref{mica}, for all values of the transmittivity t of the used beam splitter. The corresponding strategy is denoted next to each line, according to the notation used throughout this section. The gray curve represents the negativity of entanglement that can be achieved by photon subtraction from a TMSV, as found in Figure~\ref{fig:3}.}
\label{fig:4}
\end{figure}

For $\ket{\Psi_{1,0,\tilde{0},0}}$ and $\ket{\Psi_{1,0,1,0}}$, the negativity of entanglement of the output is maximum if the two measured modes are projected on coherent states. For $\ket{\Psi_{1,1,\tilde{0},\tilde{0}}}$, \ket{\Psi_{1,0,0,1}} and \ket{\Psi_{1,1,1,1}}, simple homodyne measurements - projections on eigenstates of $\hat{X}$ for $a_2$ and of $\hat{P}$ for $b_2$ - are more favourable. The states obtained by measurement are proportional to the vectors $\ket{\phi_{i,j,u,v}}$:
\begin{equation}
\begin{split}
\ket{\phi_{1,0,\tilde{0},0}}&=\ket{\psi_{1,0}}_{a_1,b_1};\\
\ket{\phi_{1,1,\tilde{0},\tilde{0}}}&=\sum_{n}(\lambda t^2)^{n}(n+2+(\lambda t^2)^2(n+1))\ket{n,n}_{a_1,b_1};\\
\ket{\phi_{1,0,0,1}}&=\sum_{n}(\lambda t)^n (n+(\lambda t)^2(n+1))\ket{n,n}_{a_1,b_1};\\
\ket{\phi_{1,0,1,0}}&=\ket{\psi_{2,\tilde{0}}}_{a_1,b_1}\\
\ket{\phi_{1,1,1,1}}&=\sum_{n}(\lambda t^2)^{n}((n+1)(n+2)+2+\\
+&(\lambda t^2)^2(((\lambda t^2)^2+2)n(n+3)+2(\lambda t^2)^2+3))\ket{n,n}_{a_1,b_1};\\
\end{split}
\label{mica}
\end{equation}
The probability of success and negativity of entanglement that can be achieved with the five measurements considered are shown in Figure~\ref{fig:4}. In comparison with a simple photon subtraction strategy, only the state $\ket{\phi_{1,0,0,1}}$
presents a better trade-off between increased entanglement and the success probability.
\begin{figure}[H]
\centering
\includegraphics[width=0.42\textwidth]{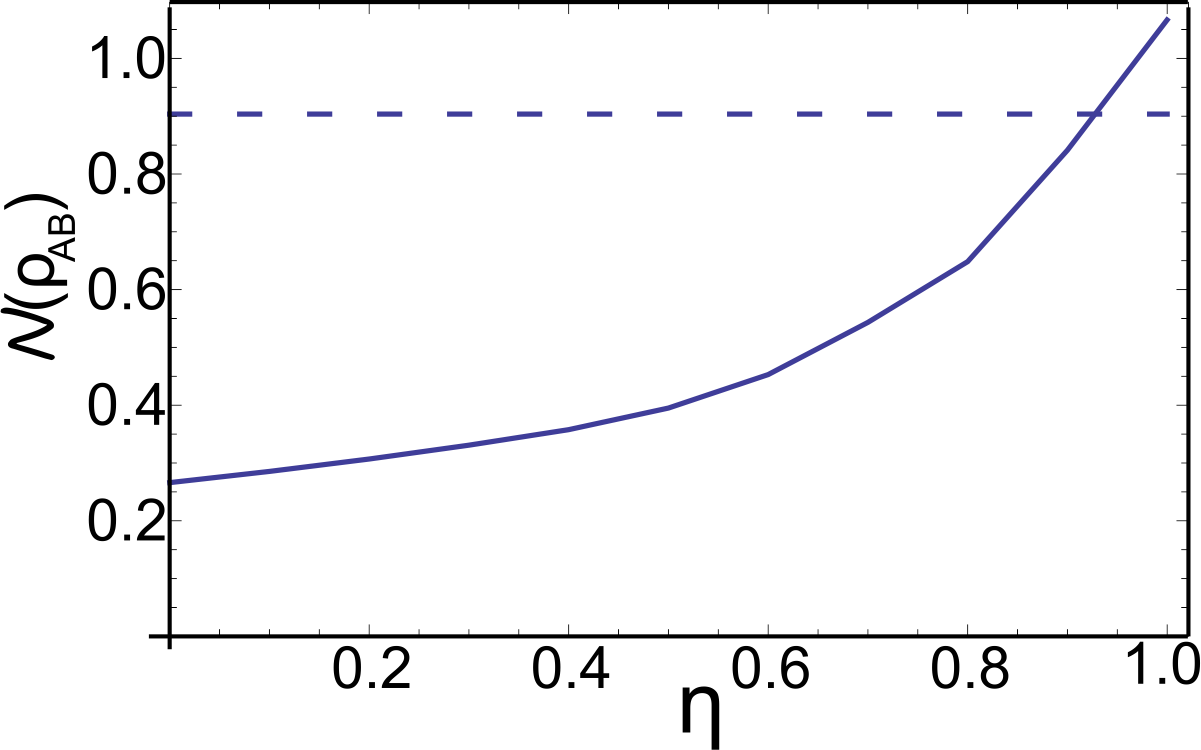}
\label{mat}
\caption{Negativity of entanglement that can be obtained by the $(1,0,0,1)$ measurement as a function of the homodyne detection efficiency $\eta$ (solid line) compared to that achieved by photon subtraction from a TMSV, with the same probability (dashed line). The transmittivity of the subtracting beam-splitter is chosen so that, for perfect detection, the difference between the values of negativity of entanglement achievable by the two compared strategies is maximum.}
\label{fig:5}
\end{figure}

It is then interesting to investigate the effect of lossy detection for this particular state. The setup is revealed to be fragile. Entanglement is increased only for detector efficiencies $\eta$ well above $0.9$, as shown in Figure~\ref{fig:5}. This could be demanding for detectors working in the pulsed regime. 

\section{Conclusions}

We have studied entanglement concentration by photon subtraction from two-mode squeezed states, within two types of setups: (i) simple photon subtraction from one two-mode squeezed state and (ii) photon subtraction from a pair of two-mode squeezed states and an implementation of the first step in the iterative distillation protocol, allowing a modified Gaussian measurement \cite{plenio2, plenio3}. Our results show that the second type of setup can, in ideal conditions, provide an advantage in terms of gained negativity of entanglement for a given success rate. However, the arrangement that allows this is highly sensitive to detector imperfection. Furthermore, a requirement for entangled states to be useful resources for CV quantum communication is that these states have a high fidelity with the closest Gaussian states \cite{ilu0}. The best-performing two-source arrangement is not symmetric and therefore does not fall within the class of operations that guarantee an evolution towards a Gaussian state. Combining our results with those in earlier work \cite{tim}, we are lead to conclude that meeting the requirements for an improvement over photon subtraction for entanglement concentration, in the presence of loss, is likely to demand and motivate further technological developments.

We thank Peter Paule for providing the code implementing Sister Celine's algorithm, Josh Nunn, Myungshik Kim, and Emilio Pisanty for useful discussions and comments. We acknowledge support from the EU IP Q-ESSENCE (248095), EPSRC (EP/J000051/1), AFOSR EOARD (FA8655-09-1-3020).  MV is supported by EPSRC through the Controlled Quantum Dynamic Center for Doctoral Training. XMJ is supported by an EU Marie-Curie Fellowship (PIIF-GA-2011-300820).

\bibliography{draftref}

\section*{Appendix}
Here, we explain the method by which the expressions of Eq. \ref{mare} can be obtained. To this end, we give the example of $\ket{\Psi_{0,0,0,0}}$.
Two initial two-mode squeezed vacuum states are represented:
\begin{equation}
\ket{\psi_{0,0}}\otimes\ket{\psi_{0,0}}=(1-\lambda^2)\sum_{n,m}\lambda^{n+m}\ket{n}_{a_1^{(0)}}\!\ket{n}_{b_1^{(0)}}\!\ket{m}_{a_2^{(0)}}\!\ket{m}_{b_2^{(0)}}
\end{equation}
Reordering the modes and applying the beam-splitter transformation to the creation operators yields:
\begin{widetext}
\begin{equation*}
\begin{split}
\ket{\Psi_{0,0,0,0}}&=\hat{B}_{a_1,a_2}\hat{B}_{b_1,b_2}(1-\lambda^2)\sum_{n,m}\lambda^{n+m}\ket{n}\ket{m}\ket{n}\ket{m}\\
&=(1-\lambda^2)\sum_{n,m}\lambda^{n+m}\hat{B}_{a_1,a_2}\frac{\hat{a}_1^{(0) n}}{\sqrt{n!}}\frac{\hat{a}_2^{(0) m}}{\sqrt{m!}}\hat{B}_{a_1,a_2}^{\dagger}\ket{00}\hat{B}_{b_1,b_2}\frac{\hat{b}_1^{(0) n}}{\sqrt{n!}}\frac{\hat{b}_2^{(0) m}}{\sqrt{m!}}\hat{B}_{b_1,b_2}^{\dagger}\ket{00}\\
&=(1-\lambda^2)\sum_{n,m}\lambda^{n+m}\frac{1}{n!m!}\left(\frac{\hat{a}_1+\hat{a}_2}{\sqrt{2}}\right)^n\left(\frac{\hat{a}_1-\hat{a}_2}{\sqrt{2}}\right)^m\left(\frac{\hat{b}_1+\hat{b}_2}{\sqrt{2}}\right)^n\left(\frac{\hat{b}_1-\hat{b}_2}{\sqrt{2}}\right)^m\ket{0000}=\\
&=(1-\lambda^2)\sum_{n,m}\lambda^{n+m}\frac{1}{2^{n+m}n!m!}\sum_{k,l,p,q}\comb{n}{k}\comb{m}{l}\comb{n}{p}\comb{m}{q}(-1)^{l+q}\hat{a}_1^{n+m-k-l}\hat{a}_2^{k+l}\hat{b}_1^{n+m-p-q}\hat{b}_2^{p+q}\ket{0000}=\\
&=(1-\lambda^2)\sum_{n,m}\lambda^{n+m}\frac{1}{2^{n+m}n!m!}\sum_{k,l,p,q}\comb{n}{k}\comb{m}{l}\comb{n}{p}\comb{m}{q}(-1)^{l+q}\sqrt{(n+m-k-l)!(k+l)!}\\
&\sqrt{(n+m-p-q)!(p+q)!}\,\ket{n+m-k-l}\ket{k+l}\ket{n+m-p-q}\ket{p+q};\\\\
\end{split}
\end{equation*}
\begin{equation}
\begin{split}
&\text{Introducing }N=n+m, K=k+l\text{ and }P=p+q, \ket{\Psi_{0,0,0,0}}=\sum_{N,K,P}\lambda^N f(N,K,P)\ket{N-K}\ket{K}\ket{N-P}\ket{P}\\
&\text{where }f(N,K,P)=\frac{1}{2^N}\left(\comb{N}{K}\comb{N}{P}\right)^{-1/2}\sum_n^N\sum_k^K\sum_p^P(-1)^{K+P-k-p}\comb{N}{n}\comb{n}{k}\comb{N-n}{K-k}\comb{n}{p}\comb{N-n}{P-p}
\end{split}
\label{equation}
\end{equation}
\end{widetext}
It can be shown that $f(N,K,P)=\delta_{K,P}$. Relabeling $i=N-K$ and $j=K$ this yields:
\begin{equation}
\begin{split}
\ket{\Psi_{0,0,0,0}}&=(1-\lambda^2)\sum_{i,j}\lambda^{i+j}\ket{i}_{a_1}\ket{i}_{b_1}\ket{j}_{a_2}\ket{j}_{b_2}\\
&=\ket{\psi_{0,0}}\otimes\ket{\psi_{0,0}}.
\end{split}
\end{equation}
This result can be confirmed by employing the Gaussian states formalism \cite{paris, review, newcv} and calculations similar to the one above can be performed for (non-Gaussian) photon-subtracted states.
Functions of the type $f(N,K,P)$ can be calculated, as they are simply sums of hypergeometrical terms. Let us denote the summand  $F(N,K,P,n,k,p)$. We have (i) the summand is 0 outside the summation intervals and (ii) ratios of the type: $$\frac{F(N+i_N,K+i_k,P+i_Q,n+i_n,k+i_k,p+i_q)}{F(N+j_N,K+j_k,P+j_Q,n+j_n,k+j_k,p+j_q},$$ where the $i$'s an the $j$'s are integers, are rational functions. $f(N,K,P)$ can be obtained by induction, after finding a recurrence relation for the sum. One way to get this is to find a recurrence for the summand $F(N,K,P,n,k,p)$, which is free of $n,k$ and $p$.\\ 
For clarity, let us look at what this means in the case of a hypergeometric term in two indices: $r$ and $s$ (the task is to sum over $s$). For some integers $I$ and $J$, one needs to find polynomials $a_{i,j}(r)$ that satisfy:
\begin{equation}
\sum_{i,j=0}^{I,J}a_{i,j}(r)F(r+j,s+i)=0.
\end{equation}
This recurrence relation can be summed over \textit{all} $s$. Because the coefficients of $F(r,s)$ in the recurrence are free of $s$ and because the summation interval can be extended, this yields a recurrence relation for the sum, $f(r)$:
\begin{equation}
\sum_{i,j=0}^{J}a_{i,j}(r)f(r+j)=0.
\end{equation}
Sister Celine's algorithm \cite{celine} is a way of finding recurrences free of a number of indices. The algorithm is described in  \cite{zeil}. The required induction is obtained computationally, using the code MultiSum \cite{multisum}, written in Mathematica \cite{mathematica}.

\end{document}